\documentclass[showpacs,preprintnumbers,amsmath,amssymb]{revtex4}
\usepackage{epsf}
\usepackage{amssymb}
\usepackage{amsfonts}
\usepackage{latexsym}
\usepackage{mathrsfs}
\usepackage{graphicx,epsf}
\def\be{\begin{equation}}
\def\ee{\end{equation}}
\def\bea{\begin{eqnarray}}
\def\eea{\end{eqnarray}}
\def\MPl{M_{\rm Pl}}

\def\D{{\Delta}}

\begin{document}
%
\title{Scale-invariant Perturbations from Chaotic Inflation}
\author{Christian T. Byrnes and David Wands}
%
%
%
\affiliation{Institute of Cosmology and Gravitation, University of
Portsmouth, Portsmouth~PO1~2EG, United Kingdom}
\date{December 7, 2005}
%
%
%
\begin{abstract}
Vacuum fluctuations in the inflaton field driving chaotic inflation with a quadratic
potential give a red spectrum of primordial density perturbations, $n=0.97$. However
angular fluctuations in an O(N)-symmetric quadratic potential have a very nearly
scale-invariant spectrum, $n=0.9998$. We investigate the possibility that these
isocurvature field perturbations could give the dominant contribution to the primordial
density perturbation after inflation.
\end{abstract}

\pacs{98.80.Cq \hfill PU-ICG-05, astro-ph/0512195}

\maketitle


\section{Introduction}

Probably the simplest model for inflation in the very early universe is an accelerated
expansion driven by an inflaton field with a quadratic potential \cite{chaotic}. In this
paper we consider the spectrum of primordial perturbations in the general case of a
quadratic potential
\begin{equation}
  \label{chaoticV}
V = \frac12 m^2 |\phi^2| \,,
\end{equation}
where $\phi$ is an $N$-component scalar field with rotational
symmetry. In this case the radial field $|\phi|$ plays the role of
the inflaton field and orthogonal fluctuations are isocurvature
perturbations during inflation as they do not effect the
energy-momentum tensor to linear order.
For example, a massive real scalar field corresponds to $N=1$,
whereas a massive complex scalar field corresponds to the case
$N=2$, and the potential energy is independent of the phase of the
field.
The model (\ref{chaoticV}) appears as a special case of inflation
driven by multiple massive fields, previously studied in models of
double inflation \cite{Polarski,Langlois} and assisted inflation
\cite{KaloperLiddle}.

Vacuum fluctuations in the inflaton field yield a lower bound on
the amplitude of the primordial density perturbation. This
spectrum deviates from exact scale-invariance, $n=1$, due to the
slope of the potential and effective mass of the inflaton field,
characterised by the dimensionless slow-roll parameters $\epsilon$
and $\eta_\sigma$ respectively. In general slow-roll inflation the
scale-dependence is given by $n-1\simeq -6\epsilon+2\eta_\sigma$
\cite{LLbook}. For a quadratic inflaton potential we have
$\epsilon=\eta_\sigma\simeq0.01$ and the perturbations have a red
spectrum with $n\simeq0.97$. Combined with the tensor-scalar
ratio, $r\simeq16\epsilon\simeq0.13$, this is marginally
compatible with current observations \cite{seljak}.

But a primordial density perturbation can also be generated from vacuum fluctuations in fields
other than the inflaton. Examples are the curvaton proposal \cite{curvaton}, modulated
reheating \cite{Kofman,DGZ} or asymmetric preheating in a nearly symmetric potential
\cite{Kolb}.
The general form for the scale-dependence of perturbations in an isocurvature field
$\chi$ during slow-roll inflation is given by $-2\epsilon+2\eta_\chi$ \cite{WBMR,BTW}.
Thus for inflation driven by the quadratic potential (\ref{chaoticV}) with
$\epsilon=\eta_\sigma=\eta_\chi$ the isocurvature perturbations have a very nearly
scale-invariant spectrum.

This paper is organised as follows. In the next section we briefly review the dynamics of
the homogeneous inflaton field driving chaotic inflation, and then calculate the
scale-dependence of the radial and angular field perturbations. We find the spectral tilt
of the isocurvature perturbations is second-order in the slow-roll approximation. In
Section III we then investigate a possible mechanism that might transfer the angular
fluctuations to the primordial density perturbation and compare this contribution with
that from inflaton perturbations. We conclude in section IV.


\section{Inflation}


We consider a scalar model with effective action
\begin{equation}
S = \int d^4x \sqrt{-g} \left[ \frac{\MPl^2}{16\pi}R - \frac12
 G^{IJ} \nabla^\mu \phi_I \nabla_\mu \phi_J - V(|\phi|) \right]
\,,
\end{equation}
where $G_{IJ}$ is the scalar field metric. We will assume the
potential, $V$, has a minimum at $\phi_K=0$ and that as it
approaches the minimum it can be described by a symmetric quadratic
potential of the form given in Eq.~(\ref{chaoticV}) where
\begin{equation}
|\phi^2| = \sum_{I,J} G^{IJ} \phi_I \phi_J \,.
\end{equation}
Henceforth we take $G_{IJ}=\delta_{IJ}$ so that the fields are canonically
normalised real fields.

We can exploit the rotational symmetry to set $\chi=\phi_J=0$
initially for all $J\geq2$ and take $\sigma=\phi_1$ to be the
inflaton field. Thus we can write
\begin{equation}
 V = \frac12 m^2 \sigma^2 \,,
\end{equation}
along the classical trajectory for any number of real scalar
fields in a symmetric quadratic potential.
This inflaton field obeys the standard equation of motion for a single scalar field
\begin{equation}
 \label{inflaton}
\ddot{\sigma}+3H\dot{\sigma}+m^2\sigma=0 \,.
\end{equation}
$\chi=0$ remains the classical solution for the spatially homogeneous background field. However
there will be quantum vacuum fluctuations in both the radial field, $\delta\sigma$, and orthogonal
fluctuations, $\delta\chi$. Radial fluctuations in the inflaton field, $\sigma$, yield adiabatic
density perturbations on large scales, while any fluctuations away from $\chi=0$ correspond to
non-adiabatic fluctuations. Non-adiabatic perturbations can in general affect the density
perturbation on large scales, but our symmetric potential has a vanishing first derivative
orthogonal to the inflaton, $V_\chi=0$, and hence the fluctuations $\delta\chi$ are strictly
isocurvature perturbations and do not couple to the adiabatic perturbations at first-order
\cite{Gordon}.

\subsection{Slow-roll parameters}

We conventionally define a dimensionless slow-roll parameter
\cite{LLbook}
\begin{equation}
\epsilon \equiv \frac{\MPl^2}{16\pi} \left( \frac{V_\sigma}{V} \right)^2 \,,
\end{equation}
where $V_\sigma\equiv \partial V/\partial\sigma$, which describes the slope of the
potential relative to the Planck scale. Two more slow-roll parameters describe the
curvature of the potential with respect to the inflaton field (down the potential) and
with respect to the orthogonal isocurvature field perturbations
\begin{eqnarray}
 \eta_\sigma &\equiv& \frac{\MPl^2}{8\pi} \frac{V_{\sigma\sigma}}{V} \,, \\
 \eta_\chi &\equiv& \frac{\MPl^2}{8\pi} \frac{V_{\chi\chi}}{V} \,.
\end{eqnarray}
For a general potential there are additional parameters at lowest order in the slow-roll
approximation proportional to $V_{\sigma\chi}$ \cite{WBMR}, but these are zero due to our
assumption of rotational symmetry.

At next order in the slow-roll expansion we also define
\begin{equation}
\xi^2_\sigma \equiv\frac{M_{Pl}^4}{(8\pi)^2} \frac{V_\sigma V_{\sigma\sigma\sigma}}{V^2}
\,, \quad \xi^2_\chi \equiv \frac{M_{Pl}^4}{(8\pi)^2} \frac{V_\sigma
V_{\sigma\chi\chi}}{V^2} \,.
\end{equation}

For the symmetric quadratic potential (\ref{chaoticV}) we have
\begin{equation}
 \label{chaoticepsilon}
\epsilon = \eta_\sigma = \eta_\chi = \frac{\MPl^2}{4\pi\sigma^2} \,, \qquad
\xi^2_\sigma=\xi^2_\chi=0\,.
\end{equation}
Note that the slow-roll parameters are thus independent of the mass,
$m$, and depend solely on the value of the inflaton field relative
to the Planck scale.

During slow-roll inflation driven by a quadratic potential the
integrated expansion, or number of e-folds from the end of inflation, is given by
\begin{equation}\label{N}
N = \int_t^{t_e} H\, dt
 \simeq \frac{2\pi (\sigma^2-\sigma_e^2)}{{\MPl}^2}
 \,.
\end{equation}
Hence the slow-roll parameter (\ref{chaoticepsilon}) is given by
\begin{equation}
\epsilon \simeq \frac{1}{2N} \left( 1+ \frac{2\pi\sigma_e^2}{N\MPl^2} \right)^{-1}
 \,,
\end{equation}
where $\epsilon_e=\MPl^2/4\pi\sigma_e^2\sim 1$ at the end of inflation. Comoving scales
comparable to our present horizon size crossed Hubble scale during inflation when
$N_{cmb}\simeq60$ \cite{LiddleLeach} and hence we have $\epsilon_{cmb}\simeq0.008$ at
that time.

\subsection{Inflaton perturbations}

Perturbations in the inflaton field are gauge-dependent, but are
most easily described in the spatially flat gauge \cite{Sasaki86}.
Inflaton perturbations with comoving wavenumber $k$ on spatially
flat slices obey the wave equation \cite{TaruyaNambu}
\begin{equation}
\ddot{\delta\sigma} + 3H \dot{\delta\sigma} + \left( \frac{k^2}{a^2} + V_{\sigma\sigma} +
\frac{8\pi}{\MPl^2a^3} \frac{d}{dt} \left( \frac{a^3\dot\sigma^2}{H} \right) \right)
\delta\sigma = 0 \,.
\end{equation}
This can be succinctly written in terms of the variable
$u=a\delta\sigma$ \cite{Mukhanov88}
\begin{equation}
u'' + \left( k^2 - \frac{z''}{z} \right) u = 0 \,,
\end{equation}
where a prime denotes derivatives with respect to conformal time $d\tau\equiv dt/a$ and
$z\equiv a\dot\sigma/H$. To leading order in the slow-roll  approximation we have
$z''/z\simeq (2+9\epsilon-3\eta_\sigma)/\tau^2$.

In order to calculate the power spectrum to leading order in the
slow-roll approximation it is sufficient to match the solution for
vacuum fluctuations on small scales, $k/a\gg H$,
\begin{equation}
 \label{vacuum}
u = \frac{1}{\sqrt{2k}} e^{-ik\tau} \,,
\end{equation}
and match it to the growing mode solution on large scales, $k/a\ll H$,
\begin{equation}
u \propto z \,,
\end{equation}
(which corresponds to a constant curvature perturbation
$H\delta\sigma/\dot\sigma$) at the time when that mode crosses the
Hubble scale: $k=aH$. We thus obtain the power spectrum for
inflaton field perturbations on large scales
\begin{equation}
 \label{Psigma}
{\cal P}_\sigma
 \equiv \frac{4\pi k^3}{(2\pi)^3} |\delta\phi^2|
 \simeq \left( \frac{\dot\sigma}{H} \right)^2 \left[ \frac{H}{\dot\sigma} \frac{H}{2\pi} \right]^2_{k=aH} \,.
\end{equation}

The scale-dependence of the spectrum (\ref{Psigma}) follows from
the time-dependence of the quantities in the square-bracket
evaluated when $k=aH$.
The spectral tilt is then given by
\begin{equation}
\Delta n_\sigma \equiv \frac{d\ln{\cal P}_\sigma}{d\ln k} \simeq -6\epsilon +2\eta_\sigma \,.
\end{equation}
Hence for a quadratic potential with $\epsilon=\eta_\sigma\simeq0.008$ we have
\begin{equation}
 \label{nsigma}
\Delta n_\sigma \simeq -0.03 \,.
\end{equation}
Note that the running of the inflaton spectral index is given by
\begin{equation}
\alpha_\sigma \equiv \frac{d\Delta n_\sigma}{d\ln k} \simeq
-24\epsilon^2+16\epsilon\eta_\sigma - 2\xi_\sigma^2 \,,
\end{equation}
where $\xi^2_\sigma=0$ for our quadratic potential and hence we
have $\alpha_\sigma\simeq -8\epsilon^2$ and hence of higher-order
in the slow-roll parameter.

\subsection{Isocurvature perturbations}

Field fluctuations, $\delta\chi$, orthogonal to the inflaton field are automatically
gauge-independent \cite{Gordon} since $\chi=0$ in the homogeneous background solution.
These are isocurvature and hence they obey the wave equation for field perturbations in an
{\em unperturbed} FRW metric:
\begin{equation}
\ddot{\delta\chi} + 3H \dot{\delta\chi} + \left( \frac{k^2}{a^2} + V_{\chi\chi} \right) \delta\chi = 0 \,.
\end{equation}
In conformal time this can be written in terms of the variable $v=a\delta\chi$ as
\begin{equation}
 \label{isov}
v'' + \left( k^2 + a^2 V_{\chi\chi} - \frac{a''}{a} \right) v = 0 \,,
\end{equation}
where to leading order in slow-roll $a''/a\simeq (2+3\epsilon)/\tau^2$ and
$m^2a^2\simeq3\epsilon/\tau^2$. Hence the equation above can be rewritten as
\begin{eqnarray}
 \label{isovnu}
v''+\left[k^2-\frac{1}{\tau^2}\left(\nu^2-\frac14\right)\right]v=0 \qquad \mathrm{where}
\qquad \nu=\frac32+\mathcal{O} (\epsilon^2) \,.
\end{eqnarray}

Note that for isocurvature perturbations in a quadratic potential the terms ${\cal
O}(\epsilon)$ in Eq.~(\ref{isov}) cancel out. Thus the equation of motion (\ref{isovnu})
to first-order in slow-roll is the same as that for a massless scalar field in de Sitter
(although the actual field is neither massless, nor in de Sitter).
%
%
$\nu$ is a constant up to second-order in slow roll parameters and hence the approximate
solution on timescales of order the Hubble time, is given by Hankel functions of order
$\nu\simeq 3/2$. Matching the general solution to the small-scale asymptotic vacuum
(\ref{vacuum}) gives
\begin{equation}\label{vsolution}
v \simeq \frac{1}{\sqrt{2k}}e^{-ik\tau}\left(1+\frac{i}{k\tau}\right).
\end{equation}

To obtain a more accurate solution over many Hubble times (without using the slow-roll
approximation), on large scales, we note that the equation of motion for isocurvature
perturbations is exactly the same  as that for the homogeneous inflaton field
(\ref{inflaton}) in the long-wavelength limit where gradient terms can be neglected.
Therefore we have $\delta\chi=A(k)\sigma $ on large scales. Matching this solution with
the growing mode of Eq.\ (\ref{vsolution}) at Hubble exit, i.e.\ at $k=aH$ or
equivalently at $k\tau=-(1+\epsilon)$ leads to
\begin{eqnarray}
\mathcal{P}_\chi(k) \equiv \frac{4\pi k^3}{(2\pi)^3} |\delta \chi|^2
 \simeq \sigma^2 \left[\frac{(1-\epsilon)H}{2\pi\sigma}\right]_{k=aH}^2
 \,.
 \end{eqnarray}
We can use the first-order slow-roll solution
\begin{equation}
\frac{H^2}{\sigma^2} \simeq \frac{4\pi}{3}\left(\frac{m}{M_{Pl}}\right)^2 \left( 1 +
\frac13 \epsilon \right) \,,
\end{equation}
at Hubble-exit, to obtain
\begin{eqnarray}
 \label{Pchi}
\mathcal{P}_\chi(k)  \simeq \frac{\sigma^2}{3\pi} \left( \frac{m}{M_{Pl}} \right)^2
 \left[1-\frac53\epsilon\right]_{k=aH} \,.
\end{eqnarray}
$\sigma^2/M_{Pl}^2\sim N/2\pi$ gives the overall time-dependence of the spectrum during
inflation.  The only scale-dependent term in this equation is the slow-roll parameter,
$\epsilon$, evaluated at Hubble-exit.

It is straightforward to calculate the spectral tilt
\begin{eqnarray}
\Delta n_\chi \equiv \frac{d \ln \mathcal{P}_\chi}{d \ln k}
 \simeq -\frac{10}{3}\epsilon^2\,.
\end{eqnarray}
Since this is second order in slow roll it is very small, $\Delta n_\chi=-0.0002$. The
running of the isocurvature spectral index is even smaller, being third-order in slow roll
\begin{equation}
\alpha_\chi\equiv\frac{d \D n_\chi}{d\ln k} \simeq -\frac{40}{3} \epsilon^3\,.
\end{equation}

In Appendix \ref{method2} we show that our calculation of the spectrum and its scale
dependence  can be confirmed using a more accurate Hankel function solution around Hubble
exit, including the second order slow-roll correction to $\nu$ in Eq.~(\ref{isovnu}).


\section{Primordial density perturbations}

We have shown that inflation driven by a quadratic potential
produces an almost scale-invariant spectrum of field perturbations
on super-Hubble scales by the end of inflation. To give rise to
the observed anisotropies in the microwave background sky and
large-scale structure in our Universe today, these field
perturbations during inflation must produce density perturbations
in the radiation dominated era after inflation. These primordial
perturbations are usefully characterised in terms of the
dimensionless density perturbation on spatially-flat
hypersurfaces. For linear perturbations we define
\begin{equation}\label{zetadefinition}
\zeta = - \frac{H\delta\rho}{\dot\rho} \,.
\end{equation}
This is equivalent to the perturbed expansion, $\delta N=H\delta
t$, up to a uniform-density hypersurface some time after
inflation.

Because we are interested in perturbations on scales very much larger than the Hubble
scale, the local expansion, $N$, is given in terms of the background solution for the
local values of the scalar fields on an initial spatially-flat hypersurface during
inflation. This is known as the ``separate universes'' approach \cite{WMLL}. Thus for one
or more scalar fields, $\phi_I$, during inflation, we have \cite{sasaki}
\begin{equation}
\zeta = \sum_I \frac{\partial N}{\partial\phi_I} \delta\phi_I \,.
\end{equation}
Lyth and Rodriguez \cite{LR} have recently pointed out that the
extension of this result to second-order also allows one to
calculate the non-Gaussianity of primordial perturbations due to
non-linear dependence of the expansion after Hubble exit on the
initial field values.
Non-Gaussianity from adiabatic field fluctuations in single-field inflation are small
(first-order in slow-roll parameters \cite{Maldacena,Komatsu}) but provide an important
constraint on non-adiabatic fluctuations in multi-field models \cite{Seery,shellard,AGW}.

The power spectrum of primordial density perturbations can thus be written as a sum of
contributions from the power spectra of individual fields
\begin{equation}
{\cal P}_\zeta = \sum_{I=1}^N \left( \frac{\partial
N}{\partial\phi_I} \right)^2 {\cal P}_{\phi_I} \,.
\end{equation}
This can then be split into a contribution from inflaton
perturbations, $\delta\sigma$, and the orthogonal isocurvature
perturbations during inflation
\begin{equation}
 \label{zetasum}
{\cal P}_\zeta = \left( \frac{\partial N}{\partial\sigma}
\right)^2 {\cal P}_{\sigma} + \sum_{I=1}^{N-1} \left(
\frac{\partial N}{\partial\chi_I} \right)^2 {\cal P}_{\chi_I} \,.
\end{equation}

The existence of a very nearly scale-invariant spectrum of
isocurvature perturbations on large scales during inflation thus allows
the possibility of producing a very nearly scale-invariant
spectrum of primordial density perturbations after inflation.
These perturbations are on wavelengths far greater than the Hubble
scale at the end of inflation and so their subsequent evolution
should be very well described by the long-wavelength approximation
in which spatial gradients can be neglected and hence is
independent of wavenumber. Thus any conversion of isocurvature to
curvature perturbations after inflation will be independent of
scale and yield a spectral tilt $n-1=\Delta n_\chi$.

\subsection{Perturbations during inflation}

Field perturbations in the inflaton field during inflation
correspond to perturbations along the background trajectory and
thus the perturbed expansion can be calculated anytime after
Hubble-exit, to give
\begin{equation}
 {\cal P}_{\zeta,{\rm inf}} = \left( \frac{H}{\dot\sigma} \right)^2 {\cal P}_\sigma \,.
\end{equation}
Using Eq.~(\ref{Psigma}) and the slow-roll approximation (\ref{N}), we have
\begin{equation}
 \label{Pzetainf}
 {\cal P}_{\zeta,{\rm inf}} \simeq \frac{16\pi}{3}  \left[ \frac{m\sigma^2}{M_{Pl}^3} \right]^2_{k=aH}
 \simeq \frac{4}{3\pi} \left( \frac{m}{M_{Pl}} \right)^2 \left[ N + \frac{2\pi\sigma_e^2}{M_{Pl}^2} \right]^2_{k=aH}
\end{equation}
Given the WMAP normalisation, $\mathcal{P}_\zeta\simeq2\times10^{-9}$ \cite{WMAP}, then
for $m\simeq 10^{-6} M_{\rm Pl}$ fluctuations in the inflaton field produce primordial
density perturbations of the observed magnitude. On the other hand for $m<10^{-6} M_{\rm
Pl}$ the inflaton perturbations are below the observational bound and we would require an
additional contribution from isocurvature perturbations to generate the primordial
density perturbations.

We also see that primordial density perturbations due to inflaton
fluctuations will have the same spectral tilt (\ref{nsigma}) as the
inflaton perturbations on large scales during inflation,
conventionally written as $n=1+\Delta n_\sigma\simeq 0.97$.

It has been shown that the non-Gaussianity from inflaton
fluctuations during slow-roll will be negligible with
non-linearity parameter $f_{NL}\simeq -5\epsilon/3$
\cite{Maldacena}.


\subsection{Perturbations from instant preheating}

Perturbations in the orthogonal fields do not affect the expansion
history during inflation due to our assumption that the potential,
$V$, is symmetric under rotations. But these isocurvature
perturbations during inflation can give an additional contribution to the density
perturbation in the radiation era if the symmetry of the self-interaction
potential (\ref{chaoticV}) is broken by other interactions of the field. In
particular they will produce density perturbations if the local
expansion, $N$, is sensitive to the initial isocurvature perturbations, $\partial
N/\partial\chi\neq0$ in Eq.~(\ref{zetasum}).

We will consider a model of instant preheating at the end of
inflation where the inflaton field's energy density can be
transferred to a preheat field during the first oscillation of the
inflaton field \cite{felder}. This simplifies the calculation of
particle production as a function of the initial field values,
making it possible to estimate the effect on preheating of the
isocurvature perturbations during inflation. It may also be the
case that instant preheating maximises the effect of the
isocurvature perturbations which might be dissipated if reheating
took place over many oscillations.

In this scenario the preheat particles $\psi$ are created through an interaction term
$g^2\phi^2\psi^2/2$, during a brief period around $\phi=0$, when their effective mass
vanishes. The energy density in the $\chi$-field is then ``fattened" by the coupling to
the inflaton field as $\phi$ rolls back up the potential because of their effective mass,
$m_\psi^2=g^2\phi^2$. Through a Yukawa interaction $\lambda\psi\omega\overline{\omega}$
the preheat field can then decay into $\omega$ particles, decaying most rapidly when the
$\psi$ particles reach their maximum effective mass, as the inflaton field reaches its
maximum. Depending on the coupling constants this process may be so efficient that all
further decay of the inflaton field can be neglected \cite{reheating,felder}.

The efficiency of instant preheating is sensitive to the phase of a complex scalar field
if the real and imaginary components ($\phi_1$ and $\phi_2$) have different couplings to
the preheat field. We consider an interaction term
\begin{equation}
 \label{interaction}
\mathcal{L}_{\textrm{int}}=-\frac12(g_1^2\phi_1^2+g_2^2\phi_2^2)\psi^2\,,
\end{equation}
which explicitly breaks the rotational symmetry of the inflationary potential (\ref{chaoticV}).

Particle creation in the preheat field first occurs when the adiabatic condition fails,
$|\dot{m}_\psi|=m_\psi^2$, shortly before the inflaton first passes through its
minimum. Denoting this time with a ``$*$'', so that e.g.\,$\sigma_*=\sigma(t_*)$, the
adiabatic condition is broken when
\begin{equation}
|\dot{\sigma}_*|=|\sigma_*|^{2}{\tilde{g}}
\end{equation}
where the effective coupling constant is given by
\begin{equation}
\tilde{g}^2(\theta) \equiv g_1^2\cos^2\theta+g_2^2\sin^2\theta
\end{equation}
and $\phi_1=\sigma\cos\theta$ and $\phi_2=\sigma\sin\theta$.
In practice $|\dot\sigma_*|\sim mM_{\rm Pl}$ and is independent of $\tilde{g}$, and hence
independent of the phase $\theta$, for $m\ll \tilde{g}M_{\rm Pl}$.

The time interval during which particle creation takes place is
\begin{equation}
\Delta t_*\sim\frac{\sigma_*}{|\dot{\sigma}_*|}
=|\dot{\sigma}_*|^{-1/2}{\tilde{g}}^{-1/2}\,.
\end{equation}
After the inflaton field has passed through the origin for the first time the occupation
number of the $\psi$ field with wavenumber $k$ is \cite{felder}
\begin{eqnarray}
n_k=\exp[-\pi(k\Delta t_*)^2]\,.
\end{eqnarray}
We integrate $n_k$ to give the total number density of
$\chi$-particles produced
\begin{eqnarray}
n_\chi=\frac1{(2\pi)^3}\int^\infty_0d^3k\,n_k =(2\pi\Delta
t_*)^{-3}
\,.
\end{eqnarray}

The ratio of the energy density of the $\chi$-field to that of the
inflaton field at the point when the inflaton would be at its
maximum after the first explosive creation of $\psi$ particles is
$\rho_\psi/\rho_\sigma\sim \tilde{g}^{5/2}(10^{-6}M_{\rm
Pl}/m)^{1/2}$ \cite{felder}. Thus if $\tilde{g}>1$ we see that
most of the inflaton's energy density is transferred to the
preheat field during the first oscillation for $m\sim
10^{-6}M_{\rm Pl}$. This constraint on $\tilde{g}$ is dependent on
$m$ and the energy transfer is more efficient for $m\ll
10^{-6}M_{\rm Pl}$, which would correspond to the inflaton
perturbation (\ref{Pzetainf}) being smaller than the observed
primordial perturbation.

Assuming the decay products are massive, non-relativistic particles, then the expansion
from an initial spatially-flat hypersurface to a final uniform-density hypersurface is
given by $N \propto (1/3)\log n_\psi$. Hence the linear metric perturbation
(\ref{zetadefinition}) is given by $\zeta=(1/3)\delta n_\psi/n_\psi$. Thus if the
effective coupling, $\tilde{g}$, is dependent upon the phase of the complex field we have
\begin{equation}\label{preheatzeta}
\zeta\simeq \frac12\frac{\partial\log \tilde{g}}{\partial
\theta}\delta\theta = \frac{F(R,\tan\theta)}{2}
\frac{\delta\chi}{\sigma} \,,
\end{equation}
where
\begin{equation}
F(R,\tan\theta) \equiv \frac{\tan\theta(1-R^2)}{R^2+\tan^2\theta}
 \quad \mathrm{and} \quad
R \equiv \frac{g_1}{g_2}\,.
\end{equation}
We only need to consider the case $R<1$ and $0<\theta<\pi/2$, due to the symmetries of
this model. The power spectrum is then given from Eqs~(\ref{Pchi}) and (\ref{preheatzeta}) by
\begin{equation}
\label{Pzetaiso}
\mathcal{P}_{\zeta,\textrm{iso}}
\simeq \frac{F^2(R,\tan\theta)}{12\pi} \left( \frac{m}{M_{Pl}}
\right)^2 \,.
\end{equation}

Comparing this with Eq.~(\ref{Pzetainf}) we see that either
$\mathcal{P}_{\zeta,\textrm{iso}}$ or $\mathcal{P}_{\zeta,\textrm{inf}}$ can dominate the
primordial density perturbation depending on the values of $R$ and $\theta$. The
asymmetric preheating creates the dominant primordial perturbation if
\begin{equation}
F(R,\tan\theta) > 4N \simeq 240 \,.
\end{equation}
We require $R<1/480$ for the isocurvature perturbations to dominate in any range of
$\theta$. If $R\ll1/480$ then the isocurvature perturbations dominate for
$240R^2<\theta<1/240$, while larger $R$ will have a smaller range of suitable $\theta$.
The range of $\theta$ where isocurvature perturbations dominate will be larger if $N<60$.
We require that the inflaton trajectory is almost exactly along $\phi_1$, the field with
the smaller coupling constant $g_1$. Although such small values for $\theta$ may, a
priori, seem unreasonably small, it may be that the interaction (\ref{interaction}) plays
a role in selecting the inflaton trajectory during inflation, but such an investigation
of the initial conditions for the inflaton are beyond the scope of this paper.

The WMAP normalisation that
$\mathcal{P}_\zeta\simeq2\times10^{-9}$ constrains the three free
parameters $m$, $R$ and $\theta$. Equation~(\ref{Pzetainf}) shows
that the primordial perturbations from the inflaton alone are of
the required size if $m \sim 10^{-6}M_{Pl}$. But if $m\ll
10^{-6}M_{Pl}$ then we require some additional source of
primordial density perturbations, and perturbations produced at
preheating from isocurvature field fluctuations are of the
required size if
\begin{equation}
F(R,\tan\theta)=200 \left( \frac{10^{-6}M_{Pl}}{m} \right) \,.
\end{equation}

Using the formalism of \cite{LR} and identifying the non-linear expansion as $N\propto
(1/3)\log(n_\psi(\theta))$, and assuming the isocurvature field perturbations themselves
are Gaussian, the non-Gaussianity of the primordial perturbation \cite{Komatsu} can be
given as
\begin{eqnarray}
f_{NL}=-\frac56\frac{N_{,\theta\theta}}{N_{,\theta}^2}
 = -\frac{5}{3(1-R^2)} \frac{(R^2-\tan^2\theta)}{\sin^2\theta} \,.
\end{eqnarray}
This is of order unity, and hence likely to be unmeasurable, for
$\theta\sim R$, but can become large for smaller $\theta$. For
example if $R=10^{-3}$, then the perturbations from asymmetric
preheating dominate in the range $0.00026<\theta<0.0039$ and
$f_{NL}$ has its maximum deviation from zero of $-24$ at
$\theta=0.00026$, but is far smaller for most of the allowed range
of $\theta$.


\section{Discussion}

We have shown that isocurvature field perturbations during chaotic inflation with  a
symmetric, quadratic potential have an almost exactly scale-invariant spectrum. We have
shown that the deviation from scale-invariance is second-order in slow-roll parameters.
The rotational symmetry of the inflationary potential leads to the scale-invariance of
the  isocurvature spectrum and keeps these field perturbations decoupled from the
adiabatic density perturbations in the inflaton field driving inflation.
But the spectrum of initial isocurvature perturbations can be transferred to the
primordial  density perturbation if interactions after inflation break the rotational
symmetry. The spectral index of the generated primordial density perturbation is then
given by $n\simeq0.9998$.

The relative importance of adiabatic inflaton fluctuations versus isocurvature field
fluctuations  in determining the primordial density perturbation in the
radiation-dominated era long after inflation depends upon the perturbation in the local
expansion \cite{sasaki}, represented by the dimensionless perturbation $\zeta$ in
Eq.~(\ref{zetasum}).
In contrast with the inflaton, the amplitude of the density
perturbations resulting from orthogonal perturbations is dependent upon the physics after inflation.
Given that the inflaton and isocurvature field perturbations on
the Hubble scale near the end of inflation are of similar
magnitude, and that the inflaton spectrum is red so grows on
larger scales, we actually require
\begin{equation}
 \left( \frac{\partial N}{\partial \chi} \right)^2 \gg \left(
  \frac{\partial N}{\partial \sigma} \right)^2 \,.
\end{equation}
for the primordial perturbations due to fluctuations in an
orthogonal field to dominate over the conventional inflaton
result.

It turns out to be rather difficult to construct a simple model
that can produce primordial density  perturbations from the
isocurvature perturbations that are larger than those from
adiabatic perturbations in the inflaton field in our chaotic
inflation model with a symmetric potential.
The expansion during slow-roll inflation is sensitive to the value of the inflaton field,
$\partial N/\partial \sigma \propto \epsilon^{-1/2}$ and hence the primordial power
spectrum is boosted by a factor $\epsilon^{-1}$ relative to the inflaton field
perturbations. If primordial density perturbations due to isocurvature field
perturbations are to dominate, then their effect on the expansion history must be boosted
by a larger factor.

In the simplest case of a single complex field, with a U(1) symmetric potential, one
could simply postulate that the efficiency of reheating at the end of inflation is an
arbitrary function of the phase, $\theta=\arctan(\phi_1/\phi_2)$. Local variations of the
inflaton decay rate, $\Gamma$, produce a primordial density perturbation \cite{DGZ}
\begin{equation}
\zeta = \frac16 \frac{\delta\Gamma}{\Gamma} \,.
\end{equation}
We require $\Gamma_{,\theta}/\Gamma\gg \epsilon^{-1}$ to produce a nearly scale-invariant
spectrum of density perturbations that dominates over the inflaton perturbations. However
as all the fields are oscillating on the same timescale, $m^{-1}$, at the end of
inflation, due to the rotational symmetry of the potential, it is not clear that one can
consistently implement the modulated reheating proposal of Refs.~\cite{DGZ,Kofman} which
assumes that the decay rate depends on the VEV of some light modulus field, which is
assumed to be approximately constant during the decay of the inflaton.

One possibility is a curvaton mechanism where the real and imaginary parts of the complex
field decayed at different rates, $\Gamma_1$ and $\Gamma_2$. However it turns out that
perturbations from a curvaton mechanism  in this model are always sub-dominant with
respect to those from inflation perturbations. The maximum expansion from the end of
inflation to the decay of the curvaton is $N=(2/3)\ln(H_e/\Gamma_1)<60$, assuming the
real part, say, decays before primordial nucleosynthesis, compared with
$N=(1/2)\ln(H_e/\Gamma_1)$ to reach the same radiation density if the imaginary part
decayed immediately at the end of inflation, $\Gamma_2\sim H_e$. Assuming $N$ is a smooth
monotonic function of the phase, $\theta$, we can estimate the perturbed expansion due to
fluctuations in the phase as
\begin{equation}
\zeta = \frac{\partial N}{\partial\theta} \delta\theta
 < \frac{15}{2\pi} \frac{\delta\chi}{\sigma} \,.
\end{equation}
{}From Eq.~(\ref{Pchi}) we thus have
\begin{equation}
{\cal P}_{\zeta,{\rm iso}}  < \frac{75}{4\pi^3} \left( \frac{m}{M_{Pl}} \right)^2
  \,.
\end{equation}
Comparing this with Eq.~(\ref{Pzetainf}) we see that the density perturbations from the
curvaton are always subdominant with respect to those from inflaton perturbations.

Instead we considered a model of preheating at the end of inflation where particle
production due to the resonant amplification of vacuum fluctuations is sensitive to the
phase of a complex scalar field. In this case we showed that the isocurvature field
fluctuations during inflation can perturb the local number density of particles produced
in the preheating field, which leads to a primordial density perturbation. Combining the
results from the inflaton perturbations (\ref{Pzetainf}) and the isocurvature
perturbations (\ref{Pzetaiso}) we obtain
\begin{equation}
 \label{Pzetaboth}
\mathcal{P}_{\zeta} = \frac{1}{12\pi} \left(\frac{m}{M_{Pl}}\right)^2
\left(16\left[N+\frac{2\pi\sigma_e^2}{M_{Pl}^2}\right]^2_{k=aH} +
\left(\frac{\tan\theta(1-R^2)}{R^2+\tan^2\theta}\right)^2 \right)
\end{equation}
We see that if $R=g_1/g_2\ll1$, corresponding to an interaction
that strongly breaks the rotational symmetry,  then the primordial
density perturbation can indeed be dominated by the nearly
scale-invariant spectrum of isocurvature field fluctuations. Even
in this model it is only possible along classical trajectories where
$\tan\theta\ll 1$.

One should bear in mind that lowering the energy scale during inflation, set by the mass
$m$, also reduces $N$, the number of e-foldings between our present horizon scale
crossing outside the Hubble scale during inflation and the end of inflation
\cite{LiddleLeach}. Throughout this paper we have taken $N\simeq 60$ when giving
numerical values which is nearly the maximum value possible. Reducing $N$ also reduces
the relative contribution of inflaton perturbations in Eq.~(\ref{Pzetaboth}).

Our model of asymmetric instant reheating is very similar in spirit to that of Kolb et al
\cite{Kolb,Kolbx2}.  Their model was based on a symmetric preheating interaction
following inflation driven by a quadratic potential with broken symmetry. The spectral
index of the isocurvature perturbations during inflation in that case can deviate from
scale-invariance due to the broken symmetry.

Ultimately, observational data will determine whether the
primordial density perturbation has a significant deviation from
scale-invariance, as predicted by adiabatic fluctuations in the
inflaton field driving chaotic inflation, or is compatible with
the very nearly scale-invariant spectrum that could be produced
from isocurvature perturbations during chaotic inflation. An
equally important ingredient in the observational constraints is
the amplitude of the primordial gravitational wave background that
is predicted. The amplitude of gravitational waves is determined
directly by the energy scale during inflation and is given by
\cite{BTW}
\begin{equation}
{\cal P}_T = \frac{32}{3\pi}\left( \frac{m}{M_{\rm Pl}} \right)^2
\left[N+\frac{2\pi\sigma_e^2}{M_{Pl}^2}\right]_{k=aH} \,.
\end{equation}

Adopting the WMAP normalisation for ${\cal P}_\zeta\simeq 2\times10^{-9}$ \cite{WMAP},
the primordial tensor-scalar ratio is given by
\begin{equation}
r \equiv \frac{{\cal P}_T}{{\cal P}_\zeta} \simeq
10^{-1}\left(\frac{m}{10^{-6} M_{Pl}}\right)^2\,.
\end{equation}
Hence we see that the tensor-scalar ratio is reduced if $m<10^{-6}M_{Pl}$. Thus if
isocurvature field fluctuations contribute to the primordial density perturbation the
gravitational wave background is smaller than if the density perturbations are due solely
to inflaton perturbations \cite{sasaki,WBMR,BTW}.

If observations rule out the relatively large tensor contribution, and deviation from
scale-invariance  of the scalar perturbations, then single-field chaotic inflation models
may be ruled out. But if the primordial perturbations originated from isocurvature field
perturbations during chaotic inflation driven by a symmetric potential, then the
primordial perturbations could be very close to a Harrison-Zel'dovich spectrum, with
$n=0.9998$, and a much smaller amplitude of gravitational waves.

\acknowledgements

CB acknowledges financial support from the EPSRC.

\appendix
\section{Hankel function method to calculate the isocurvature tilt}
\label{method2}

The equation of motion for the isocurvature perturbation is
\begin{equation}\label{seom2}
\ddot{\delta\chi}+3H\dot{\delta\chi} +\left(\frac{k^2}{a^2}+m^2\right)\delta\chi=0.
\end{equation}
This can be rewritten in conformal time and using $v=a\delta\chi$ as
\begin{equation}
v''+\left(k^2+m^2a^2-\frac{a''}{a}\right)v=0.
\end{equation}
To write the equation above in a more useful form two useful equations, that can be
derived using  the slow-roll approximation, are
$\tau=-(1/aH)(1+\epsilon+7/3\epsilon^2+\mathcal{O}(\epsilon^3))$ and
$V/3M_p^2H^2=1-\epsilon_H/3$ where
$\epsilon_H=\dot{\sigma^2}/2M_p^2H^2=\epsilon-2/3\epsilon^2+\mathcal{O}(\epsilon^3)$. It
then follows that
\begin{eqnarray}
m^2a^2=\frac{1}{\tau^2}\epsilon(3+5\epsilon+\mathcal{O}(\epsilon^2))\,,\\
\frac{a''}{a}=\frac{1}{\tau^2}(2+3\epsilon+10\epsilon^2+\mathcal{O}(\epsilon^3) \,,
\end{eqnarray}
and hence
\begin{eqnarray}
v''+\left[k^2-\frac{1}{\tau^2}\left(\nu^2-\frac14\right)\right]v = 0 \qquad
\mathrm{where} \qquad \nu=\frac32+\frac53\epsilon^2+\mathcal{O}(\epsilon^3) \,.
\end{eqnarray}

This equation can be solved in terms of Hankel functions assuming $\nu$ is constant. Note
that within a few Hubble times $\nu$ is constant up to third order in slow roll, the
solution is thus
\begin{eqnarray}
v \simeq \sqrt{-\tau}\left[c_1(k)H_\nu^{(1)}(-k\tau)+c_2(k)H_\nu^{(2)}(-k\tau)\right] \,.
\end{eqnarray}
Matching this to the small scale Klein Gordon solution $v=e^{-ik\tau}/\sqrt{2k}$ implies
$c_1(k)=e^{i(\nu+1/2)\pi/2}\sqrt{\pi}/2$ and $c_2(k)=0$. The behavior of the Hankel
function  for small argument, $z$, is
\begin{eqnarray}
H_\nu^{(1)}(z\ll1)\sim\sqrt{\frac2\pi}e^{i(\nu-1/2)\pi/2}2^{\nu-3/2}
\frac{\Gamma(\nu)}{\Gamma(3/2)} z^{-\nu}.
\end{eqnarray}
Hence the late time solution is
\begin{eqnarray}
|v| \simeq \sqrt{\frac{-\tau}{2}}(-k\tau)^{-\nu}(1+\mathcal{O}(\epsilon^2))
\end{eqnarray}
and the power spectrum is
\begin{eqnarray}
{\cal P}_\chi(k) \simeq \frac{k^3}{2\pi^2}\left(\frac{|v|}{a}\right)^2=
\frac{1}{(2\pi)^2}\frac{(-\tau)^{1-2\nu}}{a^2}k^{3-2\nu}(1+\mathcal{O}(\epsilon^2)).
\end{eqnarray}
{}From this form of the power spectrum it is simple to read off the spectral index to second-order in slow-roll parameters:
\begin{eqnarray}
\Delta n_\chi=3-2\nu \simeq -\frac{10}{3}\epsilon^2.
\end{eqnarray}

\end{document}